\documentclass[aps,prc,showpacs,nofootinbib,preprint]{revtex4}

\usepackage{epsfig}
\usepackage{graphicx,amsmath}
\newcommand\ba{\begin{eqnarray}}
\newcommand\ea{\end{eqnarray}}
\newcommand\nn{\nonumber}
\newcommand{\be}{\begin{equation}}
\newcommand{\ee}{\end{equation}}

\newcommand{\bas}{\begin{eqnarray*}}
\newcommand{\eas}{\end{eqnarray*}}

\begin{document}
\title{Experimental constraint on the
$\rho -$ meson form factors in the time--like region}
%%%
\author{A. Dbeyssi}
\affiliation{
CNRS/IN2P3, Institut de Physique Nucl\'eaire, UMR 8608, 91405 Orsay, France} 
\author{E.~Tomasi-Gustafsson}
\email[E-mail: ]{etomasi@cea.fr}
\altaffiliation{Permanent address: \it CEA,IRFU,SPhN, Saclay, 91191 Gif-sur-Yvette Cedex, France}
\affiliation{
CNRS/IN2P3, Institut de Physique Nucl\'eaire, UMR 8608, 91405 Orsay, France} 

\author{G. I. Gakh}
%%%
\affiliation{Permanent address:
\it NSC Kharkov Physical Technical Institute, 61108 Kharkov, Ukraine}
%%%
\author{C. Adamu\v s\v c\'in}
%%%
\affiliation{
\it Department of Theoretical Physics, IOP, Slovak Academy of Sciences, Bratislava, Slovakia}
%%% 
%%%

\date{\today}

\vspace{0.5cm}
\begin{abstract}
The annihilation reaction $e^++e^-\rightarrow \bar \rho+\rho $ is considered. The constraint on time-like $\rho$-meson form factors from the measurement done by the BaBar collaboration at $\sqrt{s}=10.5$ GeV is analyzed. 

\end{abstract}

\maketitle
 
\section{Introduction}

Hadron and meson electromagnetic form factors (FFs) provide important information about the 
structure and the internal dynamics of these systems. They have been object of extended experimental studies, since many decades. Presently, new facilities and detectors allow to reach high precision and to access new kinematical regions. 
A particle of spin $S$ is characterized by $2S+1$ electromagnetic FFs. The case of deuteron, which has spin one,  has been largely discussed in the literature. The individual determination of the three deuteron FFs requires the measurement of the differential cross section and at least one polarization observable, usually the tensor polarization, $t_{20}$, of the scattered deuteron in unpolarized $ed$ scattering. Data on the three deuteron FFs, charge $G_C$, quadrupole $G_Q$ and magnetic $G_M$ are available up to a momentum transfer squared $Q^2=1.9$ GeV$^2$ \cite{Abbott:2000fg}. They are best described by a model based on a six quarks hard core and a meson cloud \cite{TomasiGustafsson:2005ni}. They contradict, surprisingly, QCD predictions, even at the largest $Q^2$ value experimentally reached which corresponds to internal distances smaller than the nucleon dimension. 

The time-like (TL) region, accessible through annihilation reactions, is expected to bring a new insight to FFs. As the measurement of deuteron FFs in TL region is beyond the present experimental possibilities, it is interesting to measure the electromagnetic FFs of the $\rho $--meson, which has also spin one. The most simple reaction which contains information on TL $\rho$-meson FFs is  the annihilation of an electron--positron pair into a $\rho^+\rho^-$ pair. This question has been discussed in a previous work \cite{Adamuscin:2007dt}. Following a model independent formalism developed for spin-one particles in Ref.  \cite{Gakh:2006rj}, the differential (and total) cross sections and various polarization observables were calculated in terms of the electromagnetic FFs of the corresponding $\gamma^*\rho\rho $ current. The elements of the spin--density matrix of the $\rho-$meson were also calculated. 

The estimation of the observables was done on the basis of a simple VMD parametrization for $\rho $--meson FFs. The parameters were adjusted in order to reproduce the existing theoretical predictions in SL region \cite{MF97} where the $\rho -$meson electromagnetic FFs were calculated, both in covariant and light--front formalisms with constituent quarks. The parametrization was then analytically extended to the TL region.  Since that time, the BaBar collaboration has detected four pions identifying the $e^++e^-\to\rho^++\rho^-$ reaction \cite{Aubert:2008gm}. The results have been given in terms of helicity amplitudes.
The purpose of this work is to give the correspondence between our formalism and the helicity amplitudes and to evaluate the constraint that this unique experimental data point sets on our parameters. Relating our description of the
$\gamma^*\to \rho^+\rho^-$ vertex in terms of the electromagnetic
FFs of the $\rho -$meson with the helicity amplitudes for
this vertex, we can obtain the absolute values of FFs
moduli at the $q^2$-value where the experiment has been done.

%%%%%%%%%%%%%%%%%%%%%%%%%%%%
\section{Formalism}
%%%%%%%%%%%%%%%%%%%%%%%%%%%%

Let us consider the transition:
\begin{equation}\label{1}
\gamma^*(q)\to \rho^-(p_1)+\rho^+(p_2),
\end{equation}
where $q=p_1+p_2$, $p_1^2=p_2^2=M^2$ and M is the $\rho -$meson
mass. We consider this transition in the center of mass system (CMS) of the two $\rho
-$mesons. The 4-momenta of the  considered particles are 
$$q=(W, 0), \  \ p_1=(E, {\vec p}), \ \ p_2=(E, -{\vec p}), $$
where $W=\sqrt {q^2}$, $E ({\vec p})$ is the $\rho -$meson energy
(momentum). Let us choose the z axis along the vector ${\vec p}$,
i.e., along the $\rho^+ -$meson momentum. Then the helicity wave
functions of the particles are: 
\ba
&\epsilon_{\mu}^{(+)}&=-\frac{1}{\sqrt 2}(0, 1, i, 0),~
\epsilon_{\mu}^{(-)}=\frac{1}{\sqrt 2}(0, 1, -i, 0),~
\epsilon_{\mu}^{(0)}=(0, 0, 0, 1), \nn\\
&
U_{1\mu}^{(+)}&=-\frac{1}{\sqrt 2}(0, 1, i, 0),~
U_{1\mu}^{(-)}=\frac{1}{\sqrt 2}(0, 1, -i, 0),~
U_{1\mu}^{(0)}=\frac{1}{M}(p, 0, 0, E),~\nn\\
&
U_{2\mu}^{(+)}&=-\frac{1}{\sqrt 2}(0, -1, i, 0),~
U_{1\mu}^{(-)}=\frac{1}{\sqrt 2}(0, -1, -i, 0),~
U_{1\mu}^{(0)}=\frac{1}{M}(-p, 0, 0, E), 
\label{2}
\ea
where
$\epsilon_{\mu}^{(\lambda )}$, $U_{1\mu}^{(\lambda )}
(U_{2\mu}^{(\lambda )})$ are the wave functions of the virtual
photon and of the $\rho^- (\rho^+)$-meson with helicity $\lambda $.

%%%%%%%%%%%%%%%%%%%%%%%%%%%%%%%%%%
As the $\rho$--meson is a spin--one particle, its electromagnetic current is completely described by three FFs. Assuming P-- and C--invariance of the hadron electromagnetic interaction, this current can be written as \cite{AR77}: 
\ba
J_{\mu}&=&(p_1-p_2)_{\mu}\left[ -G_1(q^2)U_1^*\cdot U_2^*+\displaystyle\frac{G_3(q^2)}{M^2}
(U_1^*\cdot q U_2^*\cdot q-\displaystyle\frac{q^2}{2}U_1^*\cdot U_2^*)\right ]
\nn \\
&&-G_2(q^2)(U_{1\mu}^*U_2^*\cdot q-
U_{2\mu}^*U_1^*\cdot q),
\label{eq:eq5a}
\ea
where  $G_i(q^2)$ $(i=1, 2, 3)$ are the $\rho $--meson electromagnetic FFs. The FFs $G_i(q^2)$ are complex functions 
of the variable $q^2$ in the region of the TL momentum transfer ($q^2>0$). They are related to the standard $\rho $--meson 
electromagnetic FFs:  $G_C$ (charge monopole), $G_M$ (magnetic dipole) and $G_Q$ (charge quadrupole) by
\be\label{eq:eq6}
G_M=-G_2, \ G_Q=G_1+G_2+2G_3, \ \
G_C=-\displaystyle\frac{2}{3}\tau (G_2-G_3)+ \left (1-\displaystyle\frac{2}{3}\tau 
\right )G_1, \
\ \tau=\displaystyle\frac{q^2}{4M^2}\ .
\ee
or, inversely: 
\ba
G_1&&=G_Q+G_M-\frac{1}{\tau-1}[G_c-G_M-(1-\frac{2}{3}\tau)G_Q],~G_2=-G_M.\nn\\
G_3&&=\frac{1}{2(\tau-1)}[G_c-G_M-(1-\frac{2}{3}\tau)G_Q].
\label{eq:eqed}
\ea
The standard FFs have the following normalizations:
\be\label{eq:eq7}
G_C(0)=1\ , \ \  G_M(0)=\mu_{\rho}=2.14\ , \ \ G_Q(0)=-M^2Q_{\rho}=-0.79\ ,
\ee
where $\mu_{\rho}(Q_{\rho})$ is the $\rho $--meson magnetic (quadrupole) moment.

The matrix element of the $\gamma^*\to \rho^+ +\rho^-$ transition is
\begin{equation}\label{3}
M=\epsilon\cdot (p_1-p_2)[-G_1(q^2)U_1^*\cdot
U_2^*+\frac{G_3(q^2)}{M^2}(U_1^*\cdot qU_2^*\cdot
q-\frac{q^2}{2}U_1^*\cdot U_2^*)]-G_2(q^2)(\epsilon\cdot
U_1^*U_2^*\cdot q-\epsilon\cdot U_2^*U_1^*\cdot q).
\end{equation}
Let us define the following helicity amplitudes
\be
F_{\lambda_1 \lambda_2}=M_{\lambda_1 \lambda_2}^{\lambda}=
M(\epsilon\to\epsilon^{(\lambda)}, U_1\to U_1^{(\lambda_1)}, U_2\to
U_2^{(\lambda_2)}), \nn
\ee
where $\lambda_1=\lambda_{\rho^+}$,
$\lambda_2=\lambda_{\rho^-}$ and $\lambda =\lambda_{\gamma^*}$. We
have $\lambda =\lambda_1-\lambda_2$ and, therefore,
$F_{1-1}=F_{-11}=0$ since the virtual photon has spin one.
From symmetry properties it follows that $F_{-1-1}=F_{11}$ and
$F_{10}=F_{01}=F_{-10}=F_{0-1}$ and we are left with only three
independent helicity amplitudes. Let us choose
the following ones: $F_{00}$, $F_{10}$ and $F_{11}$.

The following
relation between these amplitudes and the $\rho -$meson FFs holds:
\ba
\label{4}
F_{00}&=&-\frac{\sqrt{q^2-4M^2}}{2M^2}[q^2(G_1+G_2+G_3)-2M^2G_1],  \nn\\
F_{11}&=&\sqrt{q^2-4M^2}(G_1+2\tau G_3),\nn\\
F_{10}&=&-\sqrt{\tau (q^2-4M^2)}G_2.
\label{eq:eqHA}
\ea
The value of the total cross section was evaluated in Ref. \cite{Aubert:2008gm} $\sqrt{s}=10.58$ GeV, after extrapolating beyond the experimental acceptance: $\sigma= (19.5 \pm 1.6~(stat)\pm 3.21~(syst))$ fb. 
The BaBar experiment measured also the ratio of
the moduli squared of three independent amplitudes at $\sqrt{s}=10.58$ GeV: 
\ba
{\left|{F_{00}^B}\right |^2}:{\left|{F_{10}^B}\right |^2}:{\left|{F_{11}^B}\right |^2}&=&0.51\pm 0.14~(stat)\pm 0.07~(syst):\nn\\ 
&&0.10\pm 0.04~(stat)\pm 0.01~(syst):\nn\\ 
&&0.04\pm 0.03~(stat)\pm 0.01~(syst).
\label{eq:eqfg}
\ea
where 
the following normalization was used:
\be
|F_{00}^B|^2+4|F_{10}^B|^2+2|F_{11}^B|^2=1. 
\label{eq:norm}
\ee
We are left with three unknown FFs and three independent values
(two independent ratios measured in the experiment and the normalization
condition). Thus, we can constrain the values of three FFs (moduli) at
one $q^2$ value where the experiment was done.

The total cross section of $e^++e^-\to \rho^+ +\rho^- $ \cite{Adamuscin:2007dt} can be written in terms of helicity amplitudes as:
\be
\sigma= \frac{\pi\alpha^2\beta^3}{3q^2}\frac{1}{4M^2(\tau-1)}
({\left|{F_{00}}\right |^2}+4{\left|{F_{10}}\right |^2}+2{\left|{F_{11}}\right |^2}).
\label{eq:eqfgg}
\ee

The value of the total cross section extracted from Ref. \cite{Adamuscin:2007dt} is one order of magnitude larger: $\sigma= 201$ fb. This gives an overall rescaling factor of $0.011 \pm 0.002$ GeV$^2$, which should be applied to the amplitudes extracted from parametrization \cite{Adamuscin:2007dt}. The error is calculated by propagating the experimental error on the cross section extracted from the experiment.

In our notations, Eq. (\ref{eq:norm}) reads as
\begin{equation}\label{normp}
(q^2-4M^2)[4\tau |G_2|^2+2|G_1+2\tau G_3|^2+|2\tau
(G_1+G_2+G_3)-G_1|^2]=0.011.
\end{equation}

In Ref. \cite{Adamuscin:2007dt}, the electromagnetic FFs for the $\rho$-meson were parametrized in order to reproduce the predictions from Ref. \cite{MF97} in the space-like region\footnote{Note that the square in the denominator of the expression for $G_C$ was missing in Ref. \protect\cite{Adamuscin:2007dt}, Eq. (38). }:
\ba
G_C(q^2)&=&\frac{G_C(0)(A+Bq^2)m_C^4}{(m_C^2-q^2)^2},\nn\\
G_M(q^2)&=&\frac{G_M(0)m_M^4}{(m_M^2-q^2)^2},\nn\\
G_Q(q^2)&=&\frac{G_Q(0)m_Q^4}{(m_Q^2-q^2)^2}.
\label{eq:edj}
\ea
The parameters:
$A=1$, $B$=0.33 have been fixed in order to reproduce the node of $G_C(q^2=-3$ GeV$^2$) predicted by Ref \cite{MF97}, and $m_C=1.34$ GeV, $m_M=1.42$ GeV, $m_Q=1.51$ GeV have been determined by fitting the theoretical calculation. They have the meaning of masses for the particles (mesons) carrying the interaction.

The extension of the model to TL region was made by analytical continuation, introducing an imaginary part through widths for the particles. This leads to the following parametrization:
\ba
G_C(t)&=&\displaystyle\frac{(A+Bt)m_C^4}{(m_C^2-t -im_C\Gamma_C)^2},\nn\\  
G_M(t)&=&\displaystyle\frac{G_M(0)m_M^4}{(m_M^2-t -im_M\Gamma_M)^2},\nn\\
G_Q(t)&=&\displaystyle\frac{G_Q(0)m_Q^4}{(m_Q^2-t -im_Q\Gamma_Q)^2}, 
\label{eq:eqtl}
\ea
with the following result for $\sqrt{s}=10.58$ GeV:
$|G_C|^2=1.017\cdot 10^{-4}$, $|G_Q|^2=1.167\cdot 10^{-7}$, and  $|G_M|^2=5.186\cdot 10^{-7}$. The effect of the width was illustrated in Ref. \cite{Adamuscin:2007dt} by comparing two values : $1\%$ and  $10\%$ of the corresponding mass. At large $q^2$ this effect is negligeable.

\begin{figure}
\mbox{\epsfxsize=10.cm\leavevmode \epsffile{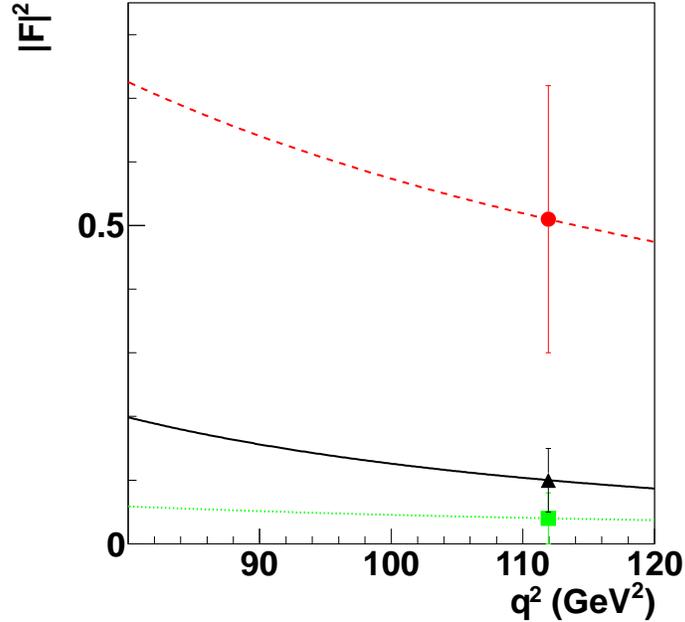}}
%\vspace*{.2 truecm}
\caption{Helicity amplitudes (moduli squared) of  $e^++e^-\to \rho^+ +\rho^- $ from Ref. \protect\cite{Aubert:2008gm}. The lines are the parametrization of the present work. Data are from Ref. \cite{Aubert:2008gm}, lines from parametrization (I): $|F_{00}|^2$ (red, circle and dashed line), $|F_{10}|^2$ (black, triangle and solid line, $|F_{11}|^2$ (green, square and dotted line).}
\label{Fig:hel}
\end{figure}

Keeping $A$ and $B$ fixed, we can re-adjust the mass parameters, such that the helicity amplitudes obtained from this parametrization coincide with those measured in the BaBar experiment (Fig. \ref{Fig:hel}). 

\begin{figure}
\mbox{\epsfxsize=10.cm\leavevmode \epsffile{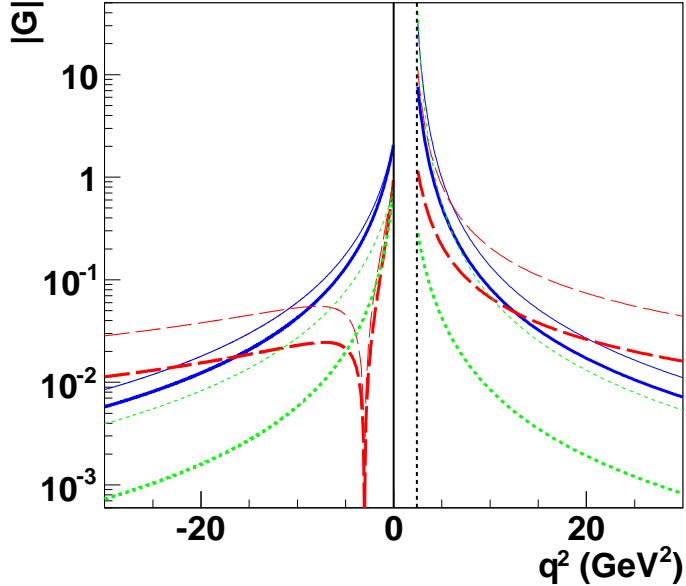}}
\caption{Absolute value of $\rho$-meson FFs $|G_i|,~i=C,Q,M$, for the parametrization (I) of the present work (thick lines) and  from Ref. \cite{Adamuscin:2007dt} (thin lines) as functions of $q^2$ in space and time-like regions: magnetic $|G_M|$ (blue, solid line), charge $|G_C|$ (red, dashed line) and quadrupole $|G_Q|$ (green, dotted line). The black, dotted line indicates the kinematical threshold for the considered reaction.}
\label{Fig:rap}
\end{figure}

The difference between the old and the present parametrization due to the experimental constraint is shown in Fig. \ref{Fig:rap}, where the moduli of the three FFs are illustrated as a function of $q^2$. The overall relative effect is small and essentially lower than an order of magnitude. In the present case we used $10\%$ width.

\begin{figure}
\mbox{\epsfxsize=10.cm\leavevmode \epsffile{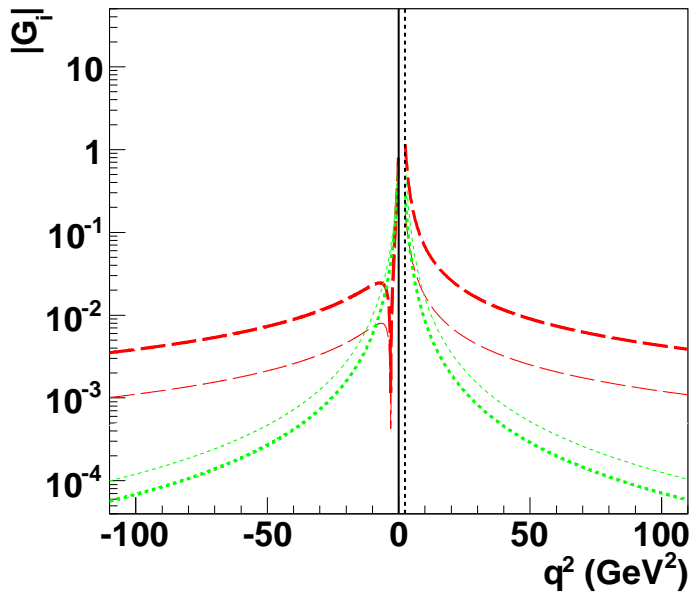}}
\caption{Absolute value of $\rho$-meson FFs $|G_i|,~i=C,Q,M$, for the parametrization (I)  (thick lines) and  (II ) (thin lines) from the present work as functions of $q^2$ in space and time-like regions: charge $|G_C|$ (red, dashed line) and quadrupole $|G_Q|$ (green, dotted line). The black, dotted line indicates the kinematical threshold for the considered reaction.}
\label{Fig:fig3}
\end{figure}

Note, that two solutions are possible for $m_C$ and $m_Q$, which can not be disentangled: the two sets ot parameters denoted as (I) and (II) in Table  \ref{tab1}, are strictly equivalent as far as the values of the amplitudes ratio and the cross section at $\sqrt{s}=10.58$ GeV are concerned, although the corresponding FFs may be different as illustrated in Fig. \ref{Fig:fig3} in an extended $q^2$-range.  
 
\begin{table}
\begin{tabular}{|c|c|c|c|}
\hline
[Ref.]&  $m_C$ [GeV]& $m_{M}$ [GeV]&$m_{Q}$ [GeV] \\
\hline
\protect\cite{Adamuscin:2007dt}  &  $1.34$ & $1.42$ &$1.51$\\
\hline
This work (I)  &  $1.05$ & $1.28$& $0.97$\\
\hline
This work (II) &  $0.77$ & $1.28$& $1.12$\\
\hline
\end{tabular}
\caption{Parameters of the model for $\rho$--meson electromagnetic FFs.}
\label{tab1}
\end{table}

%%%%%%%%%%%%%%%%%%%%%%%%%%%%%
\section{Conclusion}
%%%%%%%%%%%%%%%%%%%%%%%%%%%%%
Using the parametrization of the electromagnetic current for $\gamma^*\rho\rho $ vertex in terms
of three complex FFs, we 
compared the helicity amplitudes with the experimental value given by the BaBar collaboration.

We used a simple model for the $\rho $ meson FFs, which reproduce a calculation in SL region based on  covariant 
and light--front frameworks with constituent quarks \cite{MF97} and analytically continued to the TL region. 

In frame of VMD models, the absolute value of the amplitudes is very sensitive to the presence, the position and the width of resonances. Therefore it is interesting to compare the previous calculation with an experimental result, which is available. 

Note that the dominance of helicity conserving amplitudes in gauge theory \cite{Brodsky:1992px} implies the following ratios for the FFs of spin one bound states:
$G_C:G_M:G_Q=(1-\frac{2}{3}\tau):2:-1$. In the considered case ($\tau=46.53$), 
it  implies: $G_C:G_M:G_Q=-30:2:-1$ which is consistent with the parametrization from \cite{Adamuscin:2007dt}. However, after applying the normalization factor to the amplitudes, the following ratios have been extracted, at the corresponding $q^2$ in the space-like region: $G_C:G_M:G_Q=-63:8:-1$ for parametrization I and $G_C:G_M:G_Q=-10:5:-1$ for parametrization II. 

Therefore, as pointed out in Ref. \cite{Aubert:2008gm}, the experimental value suggests that either helicity conservation does not apply or different reaction mechanisms contribute to the $\rho$ production in the present kinematical range.

Whereas we can not draw any conclusion on the validity of the $Q^2$ dependence of our parametrization, the present comparison validates our simple approach as far as the absolute value of the cross section is concerned. Moreover, the individual helicity amplitudes can be constrained.

\section{Aknowledgments}
One of us (A.D.) acknowledges the Libanese CNRS for financial support. This work was partly supported by the GDR-PH-QCD France and by the agreement PICS No. 5419 between CNRS-IN2P3 (France) and the National Academy of Sciences of Ukraine.


\begin{thebibliography}{99}

%\cite{Abbott:2000fg}
\bibitem{Abbott:2000fg}
  D.~Abbott {\it et al.}  [JLAB t(20) Collaboration],
  %``Measurement of tensor polarization elastic electron deuteron scattering  at
  %large momentum transfer,''
  Phys.\ Rev.\ Lett.\  {\bf 84}, 5053 (2000).
%  [arXiv:nucl-ex/0001006].
  %%CITATION = PRLTA,84,5053;%%

%\cite{TomasiGustafsson:2005ni}
\bibitem{TomasiGustafsson:2005ni}
  E.~Tomasi-Gustafsson, G.~I.~Gakh and C.~Adamuscin,
  %``Two-component model for the deuteron electromagnetic structure,''
  Phys.\ Rev.\  C {\bf 73} (2006) 045204.
%  [arXiv:nucl-th/0512039].
  %%CITATION = PHRVA,C73,045204;%%



%\cite{Adamuscin:2007dt}
\bibitem{Adamuscin:2007dt}
  C.~Adamuscin, G.~I.~Gakh and E.~Tomasi-Gustafsson,
  %``Polarization effects in the reaction $e^++e^-\to \rho^+ +\rho^- $ and
  %determination of the $\rho -$ meson form factors in the time--like region,''
  Phys.\ Rev.\  C {\bf 75}, 065202 (2007).
%  [arXiv:0706.1125 [hep-ph]].
  %%CITATION = PHRVA,C75,065202;%%

%\cite{Gakh:2006rj}
\bibitem{Gakh:2006rj}
  G.~I.~Gakh, E.~Tomasi-Gustafsson, C.~Adamuscin, S.~Dubnicka and A.~Z.~Dubnickova,
  %``Polarization effects in e+ + e- --> anti-d + d and determination of time
  %like deuteron form factors,''
  Phys.\ Rev.\  C {\bf 74}, 025202 (2006)
  [arXiv:nucl-th/0604066].
  %%CITATION = PHRVA,C74,025202;%%

\bibitem{MF97}
J. P. B. C. de Melo and T. Frederico, Phys. Rev. C {\bf 55}, 2043 (1997).
%\cite{Aubert:2008gm}
\bibitem{Aubert:2008gm}
  B.~Aubert {\it et al.}  [BaBar Collaboration],
  %``Observation of e+e- ---> rho+ rho- near s**(1/2) = 10.58-GeV,''
  Phys.\ Rev.\  D {\bf 78}, 071103 (2008).
%  [arXiv:0806.3893 [hep-ex]].
  %%CITATION = PHRVA,D78,071103;%%

\bibitem{AR77} 
A. Akhiezer, M. P. Rekalo, {\it "Electrodynamics of hadrons"}, (in Russian), 
Naukova Dumka,Kiev, 1977.

%\cite{Brodsky:1992px}
\bibitem{Brodsky:1992px}
  S.~J.~Brodsky and J.~R.~Hiller,
  %``Universal Properties Of The Electromagnetic Interactions Of Spin One
  %Systems,''
  Phys.\ Rev.\  D {\bf 46}, 2141 (1992).
  %%CITATION = PHRVA,D46,2141;%%



\end{thebibliography}
\end{document}